\RequirePackage{lineno}
\documentclass[10pt]{IEEEtran}
\usepackage{amsmath,amsfonts,cite}
\usepackage{threeparttable}
\usepackage{lipsum}
\usepackage[dvips]{color}
\usepackage{graphicx}
\usepackage{lipsum}
\usepackage{algorithm,algorithmic}
\usepackage{booktabs}
\newcommand{\thickhline}{\noalign{\hrule height 1.0pt}}
\usepackage{multirow}
\ifCLASSINFOpdf
 
\else
\fi
\begin{document}
\title{Efficient Uncertainty Quantification for the Periodic Steady State of Forced and Autonomous Circuits}

\author{Zheng Zhang, Tarek A. El-Moselhy, Paolo Maffezzoni, Ibrahim (Abe) M. Elfadel, and~Luca~Daniel
\thanks{This work was supported by the MI-MIT Collaborative Program (Reference No.196F/002/707/102f/70/9374). I. M. Elfadel's work was also supported by SRC under the MEES I, MEES II, and ACE$^{4}$S programs, and by ATIC under the TwinLab program.}
\thanks{Z. Zhang, T. A. El-Moselhy and L. Daniel are with the Massachusetts Institute of Technology, Cambridge, MA, USA  (e-mail: z\_zhang@mit.edu, tmoselhy@mit.edu, luca@mit.edu).}
\thanks{P. Maffezzoni is with Dipartimento di Elettronica e Informazione, Politecnico di Milan, Milan, Italy (e-mail: pmaffezz@elet.polimi.it).}

\thanks{I.~M. Elfadel is with the Masdar Institute of Science and Technology, Abu Dhabi, United Arab Emirates (e-mail: ielfadel@masdar.ac.ae).}
}

\markboth{IEEE TRANSACTIONS ON CIRCUITS AND SYSTEMS-II: EXPRESS BRIEFS, ~Vol. ~XX, No.~XX,~XX~2013}{ZHANG \MakeLowercase{\textit{et al.}}: Efficient UQ for Periodic Steady States}

\maketitle

%

\begin{abstract}
This brief paper proposes an uncertainty quantification method for the periodic steady-state (PSS) analysis with both Gaussian and non-Gaussian variations. Our stochastic testing formulation for the PSS problem provides superior efficiency over both Monte Carlo methods and existing spectral methods. The numerical implementation of a stochastic shooting Newton solver is presented for both forced and autonomous circuits. Simulation results on some analog/RF circuits are reported to show the effectiveness of our proposed algorithms.
\end{abstract}

\begin{IEEEkeywords}
Uncertainty quantification, stochastic testing, periodic steady state, circuit simulation.
\end{IEEEkeywords}
\IEEEpeerreviewmaketitle
Ä
\section{Introduction}

\IEEEPARstart{D}ESIGNERS are interested in periodic steady-state (PSS) analysis when designing analog/RF circuits or power electronic systems. Such circuits include both forced (e.g., amplifiers, mixers, power converters) and autonomous cases (also called unforced circuits, e.g., oscillators). Popular PSS simulation algorithms include shooting Newton, finite difference, harmonic balance, and their variants~\cite{kundert:jssc99,Nastov:ieeeProc,Jacob:matrixfree,Aprille:TCAS}. 

As device sizes scale down, almost all performance metrics are influenced by manufacturing process variations. This work focuses on the uncertainty quantification (UQ) of PSS solutions under process variations. Previous perturbation techniques can be used for the sensitivity analysis of circuits with small variations~\cite{Maffezzoni:tcas2,Jacob:oscillators,Vytyaz:tcad}. However, none of them can capture the statistical information that is important for yield analysis. In order to obtain the underlying statistical information, existing mainstream circuit simulators employ Monte Carlo (MC) algorithms. MC must run a huge number of repeated simulations due to its slow convergence rate, leading to prohibitively expensive computational cost. 

Exploiting the previous development of various basis functions~\cite{PC1938,gPC2002,UQ:book}, UQ can be accelerated by stochastic spectral methods in many applications. Due to the high convergence rate, spectral methods can be much faster over MC when the number of parameters is small or medium~\cite{UQ:book}. In~\cite{Tao:2007}, polynomial chaos (PC) and harmonic balance are used to analyze forced circuits with Gaussian parameters. Generalized polynomial chaos (gPC) and stochastic Galerkin (SG) are further applied to simulate oscillators with non-Gaussian parameters~\cite{Pulch:09}. However, the resulting coupled equation causes substantial computational overhead.

This work proposes a simulator for the UQ of PSS solutions based on stochastic testing (ST) method. In~\cite{zzhang:tcad2013}, ST was proposed to simulate the DC and transient problems of transistor-level circuits. However, directly using the simulator in~\cite{zzhang:tcad2013} for PSS analysis can cause long-term integration errors. This paper employs ST to directly construct the stochastic PSS equations. We further present an efficient implementation to solve the resulting UQ equations. With our formulation, the resulting coupled PSS equation can be solved in a decoupled manner to extract the underlying statistical information, leading to substantial computational savings. This work focuses on the shooting Newton method, but extending our ideas to other types of PSS solvers is straightforward. 

\section{Background $\&$ Related Work}
\subsection{Shooting Newton Method}
Consider a general nonlinear circuit equation:
\begin{align}
\label{eq:dae}
\begin{array}{l}
 \displaystyle{\frac{{d\vec q\left( {\vec x\left( {t } \right)} \right)}}{{dt}}} + \vec f\left( {\vec x\left( {t} \right)} \right) = B\vec u\left( t \right) 
 \end{array}
\end{align}
where $\vec u(t)$ is the input signal, ${\vec x}\in \mathbb{R}^n$ denotes nodal voltages and branch currents, ${\vec q}\in \mathbb{R}^n$ and ${\vec f}\in \mathbb{R}^n$ represent the charge/flux and current/voltage terms, respectively. 

Under a periodic input $\vec u(t)$, there exists a PSS solution $\vec x(t)=\vec x(t+T)$, where the smallest scalar $T>0$ is the period known from the input. Shooting Newton method computes $y=\vec x(0)$ by solving the Boundary Value Problem (BVP)
\begin{align}
\label{det_ns_forced}
	\vec \psi (y)=\vec \phi (y,0,T )-y=0.
\end{align}
Here $\vec \phi(y,t_0,t )$ is the state transition function, which actually is the state vector $\vec x(t+t_0)$ evolving from the initial condition $\vec x(t_0)=y$. Obviously, $\vec \phi(y,0,T )=\vec x(T)$ when $y=\vec x(0)$. To compute $y$, Newton's iterations can be applied. 

For autonomous circuits, $\vec u(t)=\vec u$ is constant and $T$ is unknown, thus a phase condition must be added. For example, by fixing the $j$-th element of $\vec x(0)$, one uses the BVP
\begin{align}
\label{det_ns_unforced}
\bar \phi \left( {y ,T} \right) = \left[ {\begin{array}{*{20}c}
   {\vec \psi \left( {y ,T} \right)}  \\
   {\chi \left( {y } \right)}  \\
\end{array}} \right] = \left[ {\begin{array}{*{20}c}
   {\vec \phi \left( {y ,0,T} \right) - y }  \\
   {y_{j}  - \lambda }  \\
\end{array}} \right] = 0
\end{align}
to compute $y=\vec x (0)$ and $T$. Here $y_{j}$ is the $j$-th element of $y$, and $\lambda$ is a properly selected scalar constant. 

More details on shooting Newton can be found in~\cite{kundert:jssc99,Nastov:ieeeProc,Jacob:matrixfree,Aprille:TCAS}.

\subsection{Generalized Polynomial Chaos (gPC) Expansion}
When uncertainties are involved, (\ref{eq:dae}) is modified to
\begin{equation}
\label{eq:sdae}
\begin{array}{l}
 \displaystyle{\frac{{d\vec q( {\vec x( {t,\vec \xi } ),\vec \xi } )}}{{dt}}} + \vec f( {\vec x( {t,\vec \xi }),\vec \xi } ) = B\vec u( t ).
 \end{array}
\end{equation}
Here ${\vec \xi}\in$$\mathbb{R}^d$ denotes $d$ independent Gaussian and/or non-Gaussian parameters in the stochastic space $\Omega$. If ${\vec x( {t,\vec \xi } )}$ is a second-order stochastic process, it can be approximated by a truncated gPC expansion:
\begin{align}
\label{gpc:deno}
	\vec x ( {t,\vec \xi } )\approx \tilde x (t,\vec \xi)=\sum\limits_{k = 1}^K {\hat x^k }(t) H_{k } ( {\vec \xi } ),
\end{align}
where $\hat x^k(t)$ is a coefficient vector, and $H_{k } ( {\vec \xi } )$ is a multivariate gPC basis function satisfying
\begin{equation}
\left\langle {H_k ( {\vec \xi } ),H_j ( {\vec \xi } )} \right\rangle  = \int\limits_\Omega  \rho( {\vec \xi } )H_k ( {\vec \xi } )H_j ( {\vec \xi } )d\vec \xi   = \delta _{k,j}. 
\end{equation}
Here $\rho( {\vec \xi } )$ is the joint probability density function (PDF) of $\vec \xi$. Details on gPC basis construction can be found in~\cite{gPC2002,UQ:book}. If the highest total order of the gPC bases in (\ref{gpc:deno}) is $p$, then the total number of basis functions is
\begin{equation}
\label{Kvalue}
K = \left( \begin{array}{l}
 p + d \\ 
 \;\;p \\ 
 \end{array} \right) = \frac{{(p + d)!}}{{p!d!}},
\end{equation}
where $d$ is the number of random parameters.

Several spectral methods can be applied to solve (\ref{eq:sdae}), including the stochastic collocation (SC), stochastic Galerkin (SG) and stochastic testing (ST) methods. ST uses a collocation testing scheme to set up a coupled equation with fewer nodes than the mainstream SC and then directly computes the gPC coefficients by an intrusive solver. It is called Stochastic Testing to distinguish it from the non-intrusive SC methods that compute the gPC coefficients indirectly. Detailed comparisons of ST, SC and SG can be found in~\cite{zzhang:tcad2013}.

Stochastic Galerkin (SG) was employed in~\cite{Pulch:09} for the UQ of oscillators. The SG-based solver consists of three steps: 1) similar to~\cite{Vytyaz:tcad}, the time axis is scaled such that the scaled waveforms have a constant oscillation period; 2) the state vector and period are approximated by a truncated gPC expansion, and a coupled deterministic DAE is set up by Galerkin testing; 3) $K$ phase conditions are added, and the gPC coefficients are computed by Newton's iterations. The shortcoming of~\cite{Pulch:09} is the computational cost growing cubically w.r.t. the number of basis function $K$.

\section{Proposed ST-based PSS Solver}
\textbf{Notation.} Let ${\cal H}  $$= $$\{ {H_1 ( {\vec \xi } ), \cdots ,H_K ( {\vec \xi } )} \}$ represent the gPC basis functions, and $\hat{\textbf{w}}=[\hat w^1; \cdots; \hat w^K]$ denote the collection of gPC coefficients, we define an operator:
\small
\begin{align}	
\mathbb{M}({\cal H} ,\hat{\textbf{w}},\vec \xi ): = \tilde{w}(\vec \xi)=\sum\limits_{k = 1}^K {\hat w^k H_k (\vec \xi )} \nonumber
\end{align} \normalsize
which converts $\hat{\textbf{w}}$ to a gPC approximation $\tilde{w}(\vec \xi)$. Given a set of testing nodes ${\cal S}=\{\vec \xi_1,\cdots,\vec \xi_K \}$, $\textbf{V}\in \mathbb{R}^{K\times K}$ denotes a Vandermonde-like matrix, the $(i,j)$ element of which is
\begin{equation}
\textbf{V}_{i,j} = {H_j (\vec \xi _i )}, \; {\rm for} \; 1\leq i,j\leq K.
\end{equation}
Finally we denote $\textbf{W}_n=\textbf{V} \otimes \textbf{I}_n$, where $\otimes$ is the Kronecker product operator and $\textbf{I}_n$ is an identity matrix of size $n$.

\subsection{Formulation for Forced Circuits}
For a forced circuit, we directly perform UQ based on the following coupled DAE formed by ST
\begin{align}
\label{ST:forced}
\frac{{dQ(\hat{\textbf{x}}(t))}}{{dt}} + F(\hat{\textbf{x}}(t),t) = 0.
\end{align}
Here $\hat{\textbf{x}}(t)=[\hat x^1(t);\cdots;\hat x^K(t)]$$\in $$\mathbb{R}^{nK}$ collects the gPC coefficients of $\vec x(t,\xi)$. Let $\tilde x(t,\vec \xi):=\mathbb{M}({\cal H},\hat{\textbf{x}}(t),\vec \xi )$ and
\begin{align}
\begin{array}{l}
\vec q_k (\hat{\textbf{x}}(t)) = \vec q(\tilde x (t,\vec \xi_k),\vec \xi _k ),\\
\vec f_k (\hat{\textbf{x}}(t),t) = \vec f (\tilde x (t,\vec \xi _k),\vec \xi _k )-B \vec u(t),
 \end{array} \nonumber
\end{align}
then (\ref{ST:forced}) is obtained by the following column stacking:
\begin{align}
\begin{array}{l}
 Q(\hat{\textbf{x}}(t)) = \left[ {\vec q_1 (\hat{\textbf{x}}(t)); \cdots ;\vec q_K (\hat{\textbf{x}}(t))} \right], \\ 
 F(\hat{\textbf{x}}(t),t) = \left[ {\vec f_1 (\hat{\textbf{x}}(t),t); \cdots ;\vec f_K (\hat{\textbf{x}}(t),t)} \right].
 \end{array} \nonumber
\end{align}
ST first generates a set of multivariate quadrature nodes, then only a small port of those nodes are selected as testing nodes such that $\textbf{V}$ is invertible and well conditioned~\cite{zzhang:tcad2013}. 

The state vector $\vec x(t,\vec \xi)$ is periodic for any $\vec \xi\in \Omega$ if and only if $\hat{\textbf{x}}(t)$ is periodic. Therefore, we have
\begin{align}
\label{st_pss_forced}
 \textbf{g} (\hat{\textbf{y}}) =  \Phi (\hat{\textbf{y}},0,T) - \hat{\textbf{y}} = 0.
\end{align}
In this equation, $\hat{\textbf{y}}=\hat{\textbf{x}}(0)$, and $\Phi(\hat{\textbf{y}},0,T)$ is the state transition function of (\ref{ST:forced}).

\subsection{Formulation for Autonomous Circuits}
For unforced cases, we cannot directly use (\ref{ST:forced}) for PSS analysis since no PSS solution exists. Instead, we modify (\ref{ST:forced}) by scaling the time axis as done in~\cite{Vytyaz:tcad}. Let $T_0$ be the oscillation period for the nominal case, we write $T(\vec \xi)$ as
\begin{align}
T(\vec \xi ) = T_0 a(\vec \xi ) \approx T_0 {\mathbb M}({\cal H},\hat {\textbf{a}},\vec \xi ) \nonumber
\end{align}
where $\hat {\textbf{a}}$$=$$[\hat {a}^1;\cdots;\hat {a}^K]$ collects the gPC coefficients of $a (\vec \xi)$. Define a new time variable $\tau$ such that
\begin{align}
t = a (\vec \xi ) \tau \approx {\mathbb M}({\cal H},\hat {\textbf{a}},\vec \xi ) \tau, \nonumber
\end{align}
then $\vec z(\tau,\vec{\xi})=\vec x (t,\vec \xi)$ solves the following DAE:
\begin{equation}
\label{eq:sdae_scale}
\begin{array}{l}
 \displaystyle{\frac{{d\vec q( {\vec z( {\tau,\vec \xi } ),\vec \xi } )}}{{d\tau}}} + a (\vec \xi ) \vec f( {\vec z( {\tau,\vec \xi }),\vec \xi } ) = a (\vec \xi ) B\vec u.
 \end{array}
\end{equation}
Replacing $\vec z(\tau,\vec \xi)$ and $a (\vec \xi)$ in (\ref{eq:sdae_scale}) with their gPC approximations $\tilde z(\tau,\vec \xi)$ and $\tilde a (\vec \xi)$, respectively, and enforcing the resulting residual to zero for any $\vec \xi_k\in {\cal S}$, we get
\begin{align}
\label{ST:unforced}
\frac{{dQ(\hat{\textbf{z}}(\tau))}}{{d\tau}} + F(\hat{\textbf{z}}(\tau),\hat {\textbf{a}}) = 0.
\end{align}
Here $\hat{\textbf{z}}(\tau)$$=$$[\hat z^1(\tau);\cdots;\hat z^K(\tau)]$ denotes the gPC coefficients of $\vec z(\tau,\xi)$. The nonlinear functions are decided by
\begin{align}
\begin{array}{l}
 Q(\hat{\textbf{z}}(\tau)) = \left[ {\vec q_1 (\hat{\textbf{z}}(\tau)); \cdots ;\vec q_K (\hat{\textbf{z}}(\tau))} \right] \\ 
 F(\hat{\textbf{z}}(\tau),\hat {\textbf{a}}) = \left[ {\vec f_1 (\hat{\textbf{z}}(\tau)); \cdots ;\vec f_K (\hat{\textbf{z}}(\tau))} \right],\\
 \end{array}\nonumber
\end{align}
with 
\begin{align}
\begin{array}{l}
 \vec q_k (\hat{\textbf{z}}(\tau)) = \vec q(\tilde z (\tau,\vec \xi_k),\vec \xi _k ),\\
\vec f_k (\hat{\textbf{z}}(\tau)) = \tilde a (\vec \xi_k)(\vec f (\tilde z (\tau,\vec \xi_k),\vec \xi _k )-B \vec u).
 \end{array}\nonumber
\end{align}

Let $\hat{\textbf{y}}:=[\hat{\textbf{z}}(0);\hat {\textbf{a}}]$ and fix the $j$-th component of $\vec z(0)$ at $\lambda$, then we have the following BVP equation
\small
\begin{align}
\label{st_pss_unforced}
\textbf{g} ( {\hat{\textbf{y}} } ) = \left[ {\begin{array}{*{20}c}
   {\Psi ( {\hat{\textbf{z}}(0) ,\hat{\textbf{a}}} )}  \\
   {\chi ( \hat{\textbf{z}}(0))}  \\
\end{array}} \right] = \left[ {\begin{array}{*{20}c}
   {\Phi ( {\hat{\textbf{z}}(0),0,T_0 ,\hat{\textbf{a}}} ) - \hat{\textbf{z}}(0)}  \\
   {\chi ( \hat{\textbf{z}}(0))}  \\
\end{array}} \right] = 0.
\end{align} \normalsize
Here the state transition function $\Phi ( {\hat{\textbf{z}}(0),0,T_0 ,\hat{\textbf{a}}} )$ depends on $\hat{\textbf{a}}$, and the phase constraint $\chi ( \hat{\textbf{z}}(0))=0$$\in$$ \mathbb{R}^K$ is
\begin{align}
\chi ( \hat{\textbf{z}}(0)) = \left[ \hat{\textbf{z}}_j(0) - \lambda ;\;  \hat{\textbf{z}}_{j + n}(0);\;  { \cdots ;\;}  \hat{\textbf{z}}_{j + (K - 1)n}(0)  \right] = 0. \nonumber
\end{align}

\section{Numerical Solvers}

\subsection{Coupled Solver}
To solve (\ref{st_pss_forced}) and (\ref{st_pss_unforced}), we use Newton's iteration
\begin{align}
\label{Newton}
 {\rm{solve}}\; \Delta \hat{\textbf{y}} =  \textbf{J}^{-1}(\hat{\textbf{y}}^j )\textbf{g}(\hat{\textbf{y}}^j ), \;
 {\rm{update}}\;\hat{\textbf{y}}^{j + 1} = \hat{\textbf{y}}^j  - \Delta \hat{\textbf{y}} 
\end{align}
until convergence. $\textbf{g}(\hat{\textbf{y}})$ can be evaluated by running a transient simulation of (\ref{ST:forced}) or (\ref{ST:unforced}) for one period. 
The main problem is how to evaluate the Jacobian $\textbf{J}(\hat{\textbf{y}})$ and how to solve the linear system equation in (\ref{Newton}). 

\textbf{Forced Case. }For a forced case, the Jacobian of (\ref{st_pss_forced}) is
\begin{equation}
\label{jac:forced}
\textbf{J}_{\rm forced}  = \textbf{M}_{\hat{\textbf{y}}} - \textbf{I} ,\;{\rm with}\;{\textbf M}_{\hat{\textbf y}} = \frac{{\partial \Phi \left( {\hat{\textbf y} ,0,T} \right)}}{{\partial \hat{\textbf y}}}.
\end{equation}
Here $\textbf{M}_{\hat{\textbf{y}}}$ is the Monodromy matrix of (\ref{ST:forced}), which can be obtained from linearizations along the trajectory starting from $\hat{\textbf{x}}(0)=\hat{\textbf{y}}$ to $\hat{\textbf{x}}(T)$. This step is the same as the deterministic case detailed in~\cite{Jacob:matrixfree} and thus skipped here.

\textbf{Autonomous Case.} The Jacobian of (\ref{st_pss_unforced}) reads
\begin{align}
\label{jac:unforced}
\textbf{J}_{\rm osc}  = \left[ {\begin{array}{*{20}c}
   {{\textbf J}_{11} } & {{\textbf J}_{12} }  \\
   {{\textbf J}_{21} } & 0  
\end{array}} \right].
\end{align}
Submatrix ${\textbf J}_{11}$$=$$\frac{{\partial \Psi \left( {\hat{\textbf{z}}(0),\hat{\textbf{a}}} \right)}}{{\partial \hat{\textbf{z}}(0)}}$ can be calculated in the same way of computing $\textbf{J}_{\rm forced}$; ${\textbf J}_{21}$$=$$\frac{{\partial \chi ( \hat{\textbf{z}}(0) )}}{{\partial \hat{\textbf{z}}(0)}}$ is easy to calculate since $\chi ( \hat{\textbf{z}}(0)) $ is linear w.r.t. $\hat{\textbf{z}}(0)$. Submatrix $\textbf{J}_{12}$ is
\begin{align}
\textbf{J}_{12}  = \frac{{\partial \Psi (\hat{\textbf{z}}(0),\hat{\textbf{a}})}}{{\partial \hat{\textbf{a}}}} = \frac{{\partial \Phi (\hat{\textbf{z}}(0),0,T_0 ,\hat{\textbf{a}})}}{{\partial \hat{\textbf{a}}}}=\frac{{\partial \hat{\textbf{z}}(T_0)}}{{\partial \hat{\textbf{a}}}}.
\end{align}

Let $\tau _0$$=$$0$$<$$\tau _1$$<$$\cdots$$<$$\tau _N$$=$$T_0$ be the time points and $h_k$$=$$\tau_k$$-$$\tau_{k-1}$ be the step size in the transient simulation of (\ref{ST:unforced}). We denote the discretized trajectory by $\hat{\textbf{z}}_{(k)}=\hat{\textbf{z}}(\tau_k)$. At $\tau_k$, we have
\begin{equation}
\label{integration}
Q(\hat{\textbf z}_{(k)} ) - Q(\hat{\textbf{z}}_{(k - 1)} ) = \left(\gamma _1 F(\hat{\textbf{z}}_{(k)} ,\hat{\textbf{a}}) + \gamma _2 F(\hat{\textbf{z}}_{(k - 1)} ,\hat{\textbf{a}})\right)h_k \nonumber
\end{equation}
with $\gamma_1$$=$$\gamma_2$$=$$0.5$ for Trapezoidal rule and $\gamma_1$$=$$1$, $\gamma_2$$=$$0$ for backward Euler. Taking derivatives on both sides of the above equation yields
\begin{equation}
\label{oss_chain}
\begin{array}{l}
 \frac{{\partial \hat{\textbf{z}}_{(k)} }}{{\partial \hat{\textbf{a}}}} = (\textbf{E}_k  - \gamma _1 \textbf{A}_k h_k )^{ - 1} (\textbf{E}_{k - 1}  + \gamma _2 \textbf{A}_{k - 1} h_k )\frac{{\partial \hat{\textbf{z}}_{(k - 1)} }}{{\partial \hat{\textbf{a}}}} \\ 
 \;\;\;\;\;\;\;\; + (\textbf{E}_k  - \gamma _1 \textbf{A}_k h_k )^{ - 1} h_k (\gamma _1 {\textbf P}_k  + \gamma _2 {\textbf P}_{k - 1} ) \\ 
 \end{array}
\end{equation}
with $\textbf{E}_k $$ = $$\frac{{\partial Q(\hat{\textbf{z}}_{(k)} )}}{{\partial \hat{\textbf{z}}_{(k)} }}$, $\textbf{A}_k $$ = $$\frac{{\partial F(\hat{\textbf{z}}_{(k)},\hat{\textbf{a}} )}}{{\partial \hat{\textbf{z}}_{(k)} }}$ and $\textbf{P}_k $$ =$$ \frac{{\partial F(\hat{\textbf{z}}_{(k)},\hat{\textbf{a}} )}}{{\partial \hat{\textbf{a}} }}$. Starting from $\frac{\partial \hat{\textbf{z}}_{(0)}}{\partial \hat{\textbf{a}}}$$=0$, one gets $\textbf{J}_{12}=\frac{{\partial \hat{\textbf{z}}_{(N)} }}{{\partial \hat{\textbf{a}}}}$ by iterating (\ref{oss_chain}).

Similar to the deterministic cases~\cite{kundert:jssc99,Nastov:ieeeProc,Jacob:matrixfree,Aprille:TCAS}, the Jacobian is a dense matrix due to the matrix chain operations. Therefore, solving the linear system in (\ref{Newton}) costs $O(n^3K^3)$ if a direct matrix solver is applied, similar to the cost in~\cite{Pulch:09}. 

\subsection{Decoupled Matrix Solver}
By properly choosing a transformation matrix $\textbf{P}$, Equations (\ref{st_pss_forced}) and (\ref{st_pss_unforced}) can be converted to
 \begin{align}
\label{eq:all_transform}
\textbf{P}\textbf{g}(\hat {\textbf{y}}) = \left[ {\begin{array}{*{20}c}
   {\textbf{g}_1 (\tilde y(\vec \xi _1 ))}  \\
    \vdots   \\
   {\textbf{g}_K (\tilde y(\vec \xi _K ))}  \\
\end{array}} \right],\;{\rm{with}}\;\left[ {\begin{array}{*{20}c}
   {\tilde y(\vec \xi _1 )}  \\
    \vdots   \\
   {\tilde y(\vec \xi _K) }  \\
\end{array}} \right] = \textbf{P}\hat {\textbf{y}}.
\end{align}
Consequently, the Jacobian in (\ref{Newton}) can be rewritten as
\small
\begin{align}
\label{eq:decouple_all}
\textbf{J}(\hat {\textbf{y}}) = \textbf{P}^{ - 1} \left[ {\begin{array}{*{20}c}
   {{\rm{J}}_1 } & {} & {}  \\
   {} &  \ddots  & {}  \\
   {} & {} & {{\rm{J}}_K }  \\
\end{array}} \right]\textbf{P},\;{\rm{with}}\;{\rm J}_k  = \frac{{\partial \textbf{g}_k ( {\tilde y(\vec \xi _k )} )}}{{\partial \tilde y(\vec \xi _k ) }}.
\end{align}\normalsize

\textbf{Forced Case.} We set $\textbf{P}$$=$$\textbf{W}_n$ and $\tilde y (\vec \xi_k)$$=$$\tilde x(0,\vec \xi_k)$, then
\begin{align}
\label{eq:forced_part}
\textbf{g}_k\left( {\tilde y(\vec \xi _k )} \right) = {\vec \phi \left( {\tilde x(0,\vec \xi _k ),0,T} \right) - \tilde x(0,\vec \xi _k )} = 0
\end{align}
is a shooting Newton equation for (\ref{eq:sdae}), with $\vec \xi$ fixed at $\vec \xi_k$. In (\ref{eq:forced_part}), $\tilde x(0,\vec \xi _k )$$\in$$\mathbb{R}^n$ is unknown, $\tilde x(t, \vec \xi_k)$$=$$\vec \phi \left( {\tilde x(0,\vec \xi _k ),0,t} \right)$ is the state transition function, and ${\rm J}_k$ can be formed using existing techniques~\cite{Nastov:ieeeProc}.

\textbf{Autonomous Case.} Let $\tilde y (\vec \xi_k)$$=$$[\tilde z(0,\vec \xi_k);\tilde a(\xi_k)]$, and $\textbf{P}=\textbf{W}_{n+1} \Theta $ where $\Theta$ is a proper permutation matrix, then
\begin{align}
\label{eq:osc_part}
\textbf{g}_k\left( {\tilde y(\vec \xi _k )} \right)  = \left[ {\begin{array}{*{20}c}
   {\vec \phi \left( {\tilde z(0,\vec \xi _k ),0,T_0 ,\tilde a(\vec \xi _k )} \right) - \tilde z(0,\vec \xi _k )}  \\
   {\tilde z_j (0,\vec \xi _k )}  \\
\end{array}} \right] = 0\nonumber
\end{align} \normalsize
is a shooting Newton equation for (\ref{eq:sdae_scale}), with the parameter $\vec \xi$ fixed at $\vec \xi_k$. Here $\tilde z(0,\vec \xi _k )$ and $\tilde a(\vec \xi _k )$ are the unknowns, and $\tilde z(\tau, \vec \xi_k)$$=$$\vec \phi \left( {\tilde z(0,\vec \xi _k ),0,\tau ,\tilde a(\vec \xi _k )} \right)$ is a state transition function dependent on $ a(\vec \xi)$$=$$\tilde a(\vec \xi _k )$. The small Jacobian ${\rm J}_k$ can also be formed by existing techniques~\cite{Aprille:TCAS,Vytyaz:tcad}.

\textbf{Intrusive Solver}.
We directly compute the gPC coefficients by solving (\ref{st_pss_forced}) or (\ref{st_pss_unforced}), with decoupling \textbf{inside} the Newton's iterations (\ref{Newton}). Specifically, inside each iteration, Eq. (\ref{ST:forced}) or (\ref{ST:unforced}) is first integrated for one period, and the state trajectories are converted to the gPC approximations [i.e., $\tilde x(t,\vec \xi_k)$'s in forced circuits, or $\tilde z(\tau,\vec \xi_k)$'s and $\tilde a(\vec \xi_k)$'s in unforced circuits]. Then ${\rm J}_k$'s are formed as done in existing deterministic PSS solvers~\cite{kundert:jssc99,Nastov:ieeeProc,Jacob:matrixfree,Aprille:TCAS}. Finally, based on (\ref{eq:decouple_all}) each small block is solved independently to update $\hat{\textbf{y}}^j$. Doing so allows simulating (\ref{ST:forced}) or (\ref{ST:unforced}) with flexible time stepping controls inside the intrusive transient solver~\cite{zzhang:tcad2013}, such that all components of $\hat {\textbf x}(t)$ [or $\hat {\textbf z}(\tau)$] are located on the same adaptive time grid. This allows us to directly extract the statistical information of the time-domain waveforms and other performance metrics (e.g., statistical transient power).

\textbf{Complexity.} Since $\Theta^{-1}$$=$$\Theta^T$, $\textbf{W}_n^{ - 1}$$=$$\textbf{V}^{-1}$$\otimes$$ \textbf{I}_n$ and $\textbf{V}^{-1}$ can be easily computed~\cite{zzhang:tcad2013}, the cost of decoupling in (\ref{eq:decouple_all}) is negligible. After decoupling, one can solve each small linear system equation as done in deterministic PSS solvers~\cite{kundert:jssc99,Nastov:ieeeProc,Jacob:matrixfree,Aprille:TCAS}. The total cost is $O(Kn^3)$ if a direct matrix solver is used. For large-scale circuits, one can use matrix-free iterative methods~\cite{Jacob:matrixfree} at the cost of $O(Kn^{\beta})$ where $\beta$ is normally $1.5$$\sim $$2$. This intrusive decoupled solver could be easily parallelized potentially leading to further speedup.

\section{Numerical Results}
Our algorithm was implemented in a Matlab circuit simulator. All experiments were run on a workstation with 3.3GHz 4-GB RAM.

\subsection{Low-Noise Amplifier (LNA)}
\begin{figure}[t]
	\centering
		\includegraphics[width=3.3in]{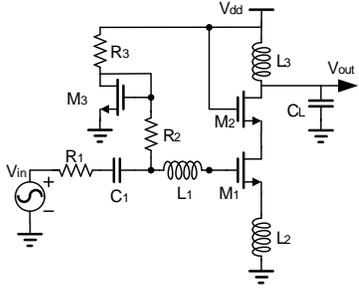} 
\caption{Schematic of the LNA.}
	\label{fig:LNA}
\end{figure}

The LNA in Fig.~\ref{fig:LNA} is used as an example of forced circuits. The ratios of the transistors are $W_1/L_1$$=$$W_2/L_2$$=$$500/0.35$ and $W_3/L_3$$=$$50/0.35$. The design parameters are: $V_{\rm dd}$$=$$1.5$ V, $R_1$$=$$50 \Omega$, $R_2$$=$$2$ k$\Omega$, $C_1$$=$$10$ pF, $C_L$$=$$0.5$ pF, $L_1$$=$$20$ nH and $L_3$$=$$7$ nH.  We introduce four random parameters. Temperature $T$$=$$300+{\cal N}(0,1600)$ K is a Gaussian variable influencing transistor threshold voltage; $R_3$$=$$1+{\cal U}(-0.1,0.1)$ k$\Omega$ and $L_2$$=$$1.4+{\cal U}(0.6,0.6)$ nH have uniform distributions; the threshold voltage under zero $V_{\rm bs}$ is $V_{\rm T}$$=$$0.4238+{\cal N}(0,0.01)$ V. The input is $V_{\rm in}=0.1{\rm sin}(4\pi\times 10^8t)$ V. 
\begin{figure}[t]
	\centering
		\includegraphics[width=3.3in]{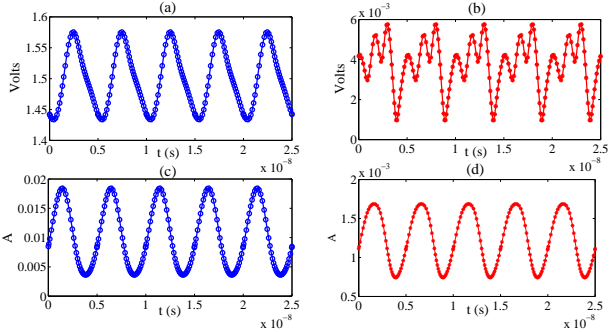} 
\caption{Periodic steady-state waveforms for the LNA. (a) $\&$ (b): mean and s.t.d of $V_{\rm out}$; (c) $\&$ (d): mean and s.t.d of $I (V_{\rm dd})$.}
	\label{fig:LNA_wave}
\end{figure}  

\begin{figure}[t]
	\centering
		\includegraphics[width=3.3in]{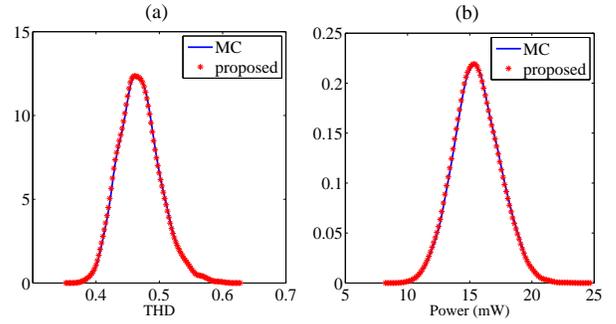}
\caption{Probability density functions. (a)THD and (b) power dissipation.}
	\label{fig:LNA_pdf}
\end{figure}

In our ST-based PSS solver, an order-$3$ gPC expansion (with $35$ basis functions) are used to represent the state variables. The computed gPC coefficients are then used to extract statistical information at a negligible cost. The means and standard deviations (s.t.d) of $V_{\rm out}$ and $I(V_{\rm dd})$ (current from $V_{\rm dd}$) are plotted in Fig.~\ref{fig:LNA_wave}. Using standard MC, $8000$ samples are required to achieve the similar level of accuracy ($<$$1\%$ relative errors for the mean and standard deviation). Fig.~\ref{fig:LNA_pdf} plots the probability density functions (PDF) of the total harmonic distortion (THD) and power consumption from our proposed PSS solver and MC, respectively. The PDFs from both methods are graphically indistinguishable. The total cost of our decoupled ST solver is $3.4$ seconds, which is $42\times$ faster over the coupled ST solver, $71\times$ faster over the SG-based PSS solver, and $220\times$ faster over MC.

\subsection{BJT Colpitts Oscillator}
\begin{figure}[t]
	\centering
		\includegraphics[width=3.3in]{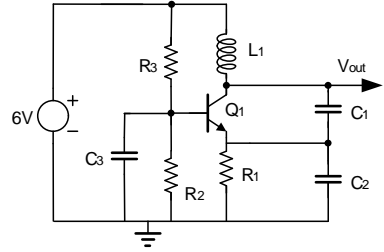} 
\caption{Schematic of the BJT Colpitts oscillator.}
	\label{fig:Colp_osc}
\end{figure}

The BJT Colpitts oscillator in Fig.~\ref{fig:Colp_osc} is a typical example of autonomous circuits. The design parameters of this circuit are $R_1$$=$$ 2.2$ k$\Omega$, $R_2$$=$$R_3$$=$$10$ k$\Omega$, $C_2$$=$$100$ pF, $C_3$$=$$0.1 \mu$F, and $\alpha$$=$$0.992$ for the BJT. The oscillation frequency is mainly determined by $L_1$, $C_1$ and $C_2$. We assume that $L_1$$=$$150+{\cal N}(0,9)$ nH and $C_1$$=$$100+{\cal U}(-10,10)$ pF are random variables with Gaussian and uniform distributions, respectively.

\begin{figure}[t]
	\centering
		\includegraphics[width=3.3in]{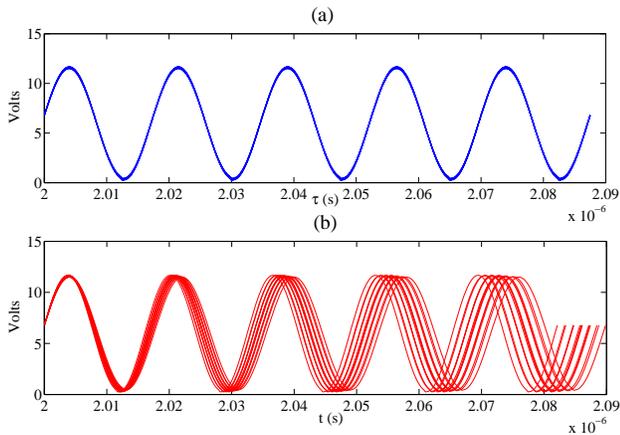} 
\caption{Realizations of $V_{\rm out}$ for the Colpitts oscillator. (a) on the scaled time axis, (b) on the original time axis.}
	\label{fig:Colp_wave}
\end{figure}

\begin{figure}[t]
	\centering
		\includegraphics[width=3.3in]{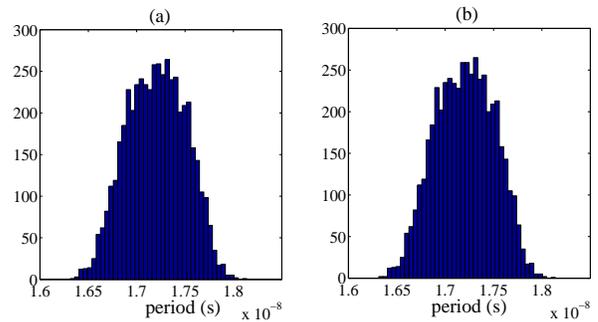} 
\caption{Distributions of the oscillation period: (a) from our proposed method, (b) from MC (5000 samples).}
	\label{fig:Colp_MC}
\end{figure}

Setting the gPC order to $3$, the results from our proposed solver and the SG-based solver~\cite{Pulch:09} are indistinguishable. Fig.~\ref{fig:Colp_wave} shows some realizations of $V_{\rm out}$ obtained by our solver. The variation looks small on the scaled time axis, but it is significant on the original time axis due to the uncertainties of the oscillation frequency. The CPU time of our decoupled ST-based solver is $4.9$ seconds, which is $2\times$ and $5\times$ faster over the coupled ST-based solver and the SG-based solver~\cite{Pulch:09}, respectively.
Finally, our solver is compared with standard MC. The computed mean and standard deviation (both in nanosecond) of the oscillation period are shown in Table~\ref{tab:MCST}. To achieve the similar level of accuracy, MC must use $5000$ samples, which is about $507\times$ slower than using our ST-based simulator. The distributions of the oscillation period from both methods are consistent, as shown in Fig.~\ref{fig:Colp_MC}.
\begin{table}[t]
	\centering
	\caption{Simulation results of the oscillation period by our proposed method and Monte Carlo.}	
	\label{tab:MCST}
	\smallskip 
  \begin{threeparttable}
	\scriptsize	
		\begin{tabular}{|c|c|c|c|c|}
	\hline
		&\multicolumn{3}{|c|}{Monte Carlo}	& Proposed \\ \hline
		$\#$ samples & 500 & 2000 & 5000 & 10\\ \thickhline		
		mean value (ns) & $17.194$ & $17.203$ & $17.205$ & $17.205$ \\  \hline		
		 s.t.d value (ns)& $2.995$ & $3.018$ & $3.026$ & $3.028$ \\
		\hline
		CPU time (s)& 252 & 1013  & 2486  & 4.9\\ \hline
	\end{tabular} 	
\end{threeparttable}	
\end{table}

\subsection{Accuracy and Efficiency}
\begin{figure}[t]
	\centering
		\includegraphics[width=3.3in]{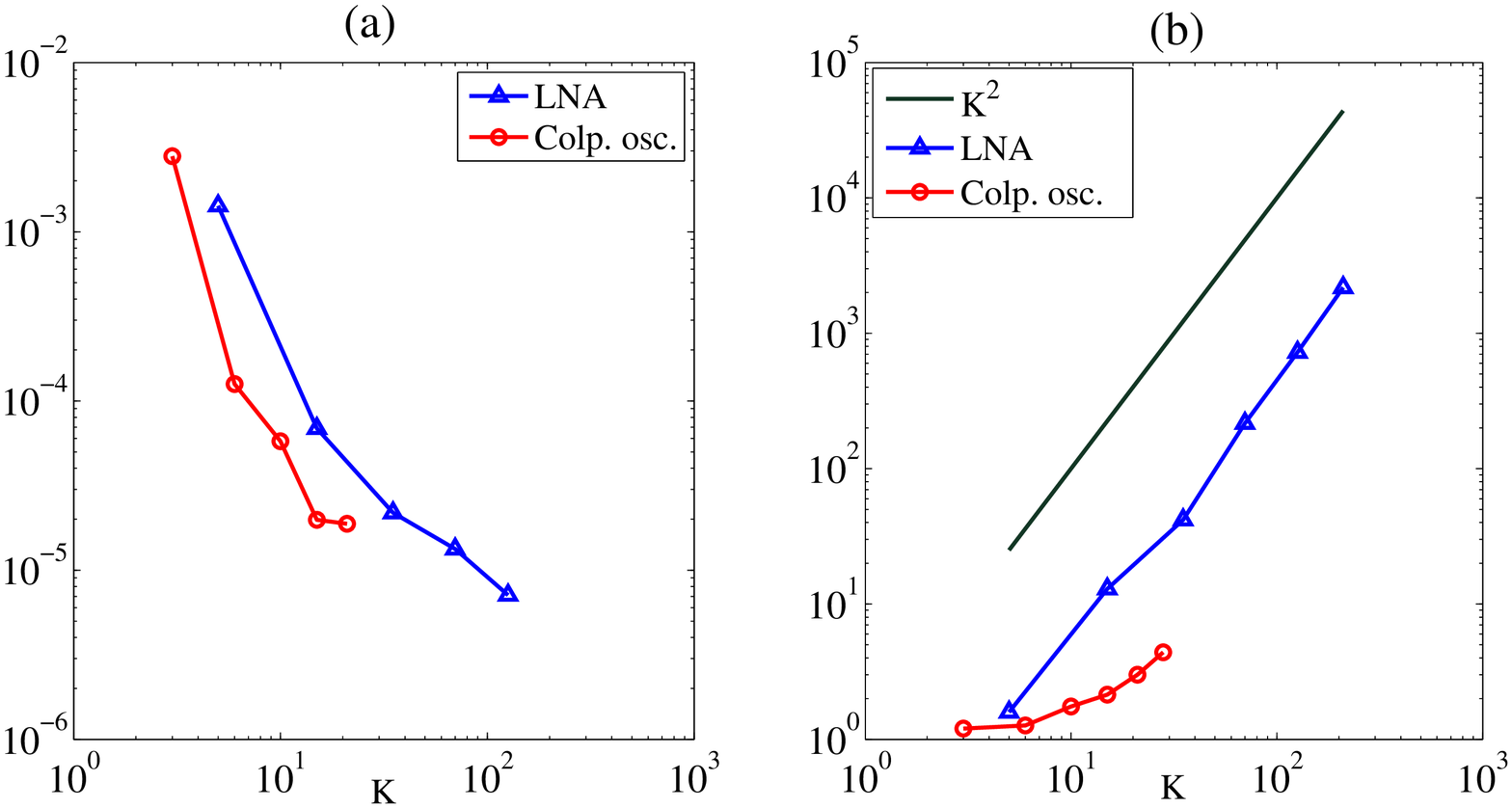} 
\caption{(a) Relative error of our solver. (b) Speedup factor caused by decoupling.}
	\label{fig:convg}
\end{figure} 
We increased the gPC order from $1$ to $6$, and treated the results from the $6$th-order gPC expansion as the ``exact" solutions. Fig.~\ref{fig:convg} plots the relative errors of $\hat{\textbf{y}}$ and the speedup factors caused by decoupling. The errors rapidly reduce to below $10^{-4}$, and the convergence slows down when the errors approach $10^{-5}$, i.e., the threshold for the Newton's iterations which dominates the accuracy. In Fig.~\ref{fig:convg}(b), the speedup curve for the LNA has the same slope as $K^2$ on a logarithmic scale, implying an $O(K^2)$ speedup caused by decoupling. The speedup for the Colpitts oscillator is however not significant, since device evaluations dominate the total cost for this small circuit. Generally, the $O(K^2)$ speedup is more obvious for large-scale circuits.

\section{Conclusion}
This paper has proposed an intrusive periodic steady-state simulator for the uncertainty quantification of analog/RF circuits. The main advantage of our proposed method is that the Jacobian can be decoupled to accelerate numerical computations. Numerical results show that our approach obtains results consistent with Monte Carlo simulation, with $2$$\sim$$ 3$ orders of magnitude speedup. Our method is significantly faster over existing SG-based PSS solver, and the speedup factor is expected to be more significant as the circuit size and the number of basis functions increase.


\bibliographystyle{IEEEtran}
\bibliography{tcas2}

\end{document}